\documentclass[twocolumn,aps,prl]{revtex4}
\usepackage{mathrsfs}
\usepackage{amsmath}
\usepackage{amsfonts}
\usepackage{amssymb}
\usepackage{amsthm}
\usepackage{graphicx}
\begin{document}

\title{A theory for magnetic-field effects of nonmagnetic
organic semiconducting materials}
\author{X. R. Wang}
\affiliation{Physics Department, The Hong Kong University of Science
and Technology, Clear Water Bay, Hong Kong SAR, China}
\author{S. J. Xie}
\affiliation{School of Physics, National Key Laboratory of Crystal
Materials, Shandong University, Jinan, P. R. China}

\begin{abstract}
A universal mechanism for strong magnetic-field effects of
nonmagnetic organic semiconductors is presented. A weak magnetic
field (less than hundreds mT) can substantially change the charge
carrier hopping coefficient between two neighboring organic
molecules when the magnetic length is not too much longer than the
molecule-molecule separation and localization length of electronic
states involved. Under the illumination of lights or under a high 
electric field, the change of hopping coefficients leads also to the
change of polaron density so that photocurrent, photoluminescence,
electroluminescence, magnetoresistance and electrical-injection
current become sensitive to a weak magnetic field.
The present theory can not only explain all observed features,
but also provide a solid theoretical basis for the widely used
empirical fitting formulas.
\end{abstract}
%\keywords{Magnetic-field effects, nonmagnetic organic semiconductors,polaron}
\date{\today}
\maketitle
%---------------------------------------
One of the long-term\cite{merrifield,williams,wohlgenannt,
wohlgenannt1} unsolved fundamental issues in organic physics is the
mechanism behind the strong (a few per cent) responses of electrical
and optical properties of nonmagnetic organic semiconductors to a
weak magnetic field (less than hundreds mT), known as organic
magnetic-field effect (OMFE). The recent revival interest in OMFE on
the magnetoresistance, photoluminescence, photocurrent,
electroluminescence, and electrical-injection current in nonmagnetic
organic semiconductors is largely due to its importance in
fundamental science and technology applications\cite{hu}. Firstly,
there is a belief that the OMFE can be used as a powerful
experimental tool to probe useful and non-useful excited processes
of organic materials. Secondly, the OMFE can be used to develop new
multifunctional organic devices for information, sensing and energy
technologies\cite{shi}. Experiments showed that OMFE has following
surprising yet universal features. 1) The OMFE appears in vast
different organic semiconductors without any magnetic elements at
room temperature although the possible energy level shifts due to
the presence of a magnetic field are orders magnitude smaller than
the thermal energy and other energy scales. 2) The
electroluminescence, photocurrent, photoluminescence, and
electrical-injection current are very sensitive to weak magnetic
field with both positive and negative OMFE though positive OMFE (or
negative magnetoresistance (MR) in the convention terminology) at
very weak field is typically observed. 3) The OMFE can often be
fitted by two empirical formulas: $[B/(B+B_0)]^2$ and
$B^2/(B^2+B_0^2)$\cite{wohlgenannt1}, where $B$ is the applied
magnetic field. In the theoretical side, it is
known\cite{wohlgenannt,wohlgenannt1} that familiar MR mechanisms
such as Lorentz force, conventional hopping MR, electron-electron
interaction and weak localization are highly unlikely to be the
cause behind the OMFE. The current belief in the community is that
the MEFS is intimately tied to spin physics involving spin
configuration, spin correlation, and spin flip\cite{hu}. However,
there is no convincing arguments why an extremely small Zeeman
energy can beat other much larger energy scales in controlling
electron spin dynamics to generate this OMFE. Debate on whether
excitons, biexcitons, polarons, or bipolarons are the cause of the
OMFE has been going on for many years with more confusions than any
conclusion\cite{Kalinowski,Prigodin,Hu2,Robbert,Majumdar}. Even the
studies on the possible roles of the hyperfine and spin-orbital
interactions in the OMFE can hardly provide any clue to the answer
to this standing puzzle why a field of less than hundreds mT
(including contributions from hyperfine and spin-orbital
interactions) can produce such big magnetic-field effects at room
temperature. Both extensive experimental and theoretical studies so
far are suggesting a novel explanation is needed. This new MR
mechanism should explain not only all OMFE features, but also why
the similar effects does not appear in the usual inorganic
nonmagnetic semiconductors. In this report, we present such a theory
that does not explicitly rely on the electron spin degrees of
freedom. It is showed that the OMFE originates from the substantial
change of electron hopping coefficient in a magnetic field because
of narrow bandwidth of nonmagnetic organic semiconductors and large
(comparable to the magnetic length) molecule-molecule separation and
localization length.

Nonmagnetic organic semiconductors have a few distinct properties
that their inorganic counterparts do not have. Instead of covalent
chemical bonding, organic molecules in nonmagnetic organic
semiconductors are bonded by the Van der Waals force so that the
electron hopping coefficient between nearby molecules is small at
tenths eV, resulting in very narrow band instead of order of 10eV
bandwidth for their inorganic counterparts \cite{Troisi}. The
intramolecular excitons have strong binding energies of order of
eV\cite{hu}. On the other hand, the electron and hole become polaron
pair when they are located on different molecules because the
intermolecular exciton binding energy is smaller than the thermal
energy\cite{hu}. The electrical properties of an organic
semiconductor are mainly determined by the motion of polarons since
the motion of excitons does not contribute to the electric current.
The singlet excitons are, on the other hand, responsible to the
luminescence due to the spin-selection rule. A weak field should not
change much of energy levels of various excitation states so that
their populations at thermal equilibrium should not be sensitive
to a magnetic field since they are given by the Fermi-Dirac
distribution that depends only on the energy level distribution
and the temperature. Any significant change in magnetoresistance
near the quasiequilibrium state must come from the mobility change.
The question is whether a weak field of 100mT can change the
mobility of polarons in organic semiconductors.
In the usual inorganic crystals with lattice constant of order of
angstroms, the answer is no. However, we are going to argue below
that with molecule-molecule separation of tens $nm$ in organic
conjugated materials, this can indeed happen.

\begin{figure}[tbph]
\begin{center}
\includegraphics[width=0.9\columnwidth]{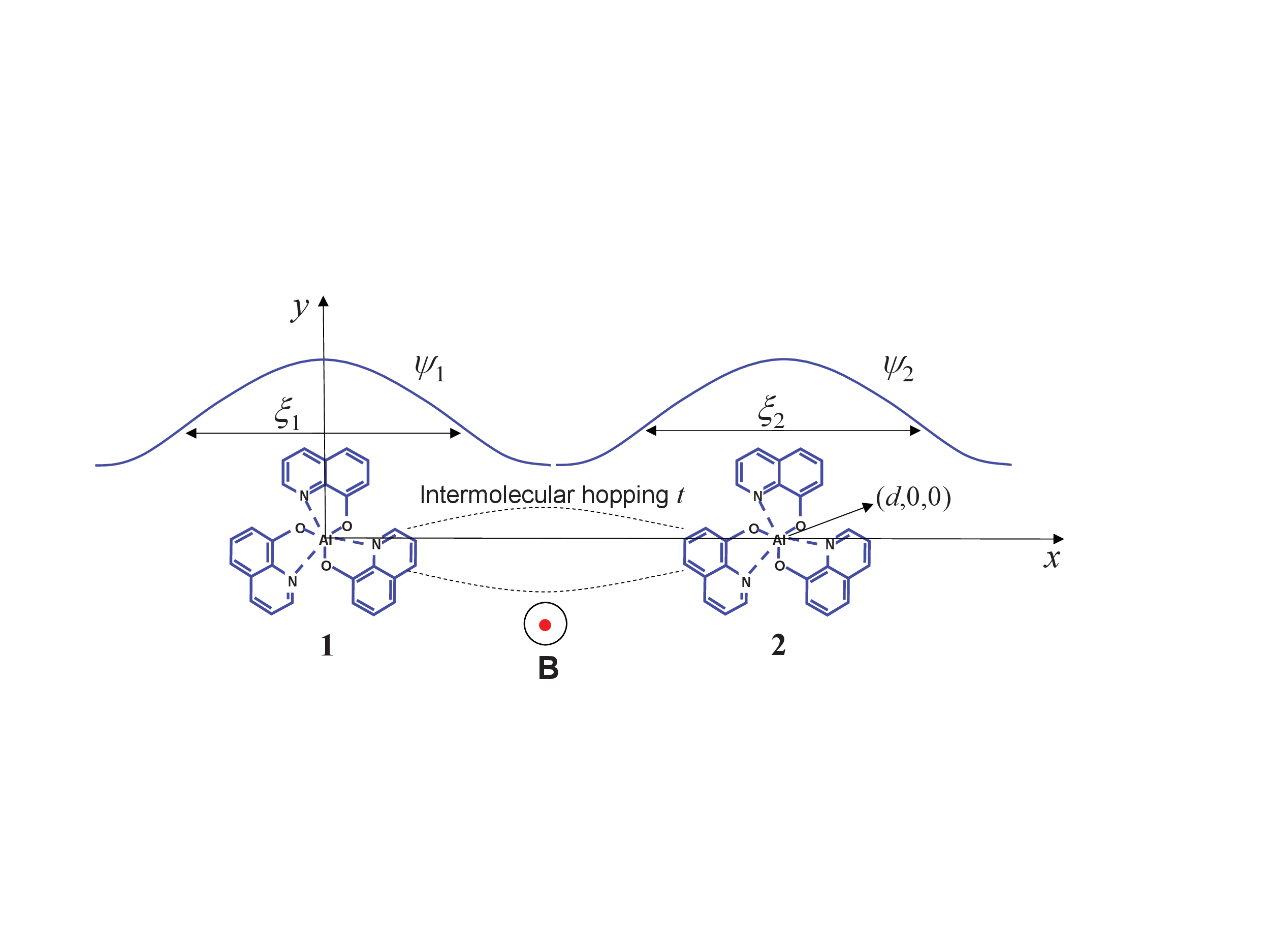}
\end{center}
\caption{Schematical draw of two organic molecules (Alq$_3$) 
separated by a distance $d$ and aligned along x-direction. 
Molecule {\bf 1} is centered at the origin, and molecule 
{\bf 2} is centered at $(d,0,0)$. The field is assumed to be 
along z-direction. $\psi_1$ and $\psi_2$ are two localized 
states with localization lengths $\xi_1$ and $\xi_2$ on 
molecules {\bf 1} and {\bf 2}, respectively. }
\label{fig1}
\end{figure}

In order to understand why a weak magnetic field can change charge
carrier (electron and hole or polaron) mobility in an organic
conjugated material, we consider a system with two molecules
separated by a distance $d$ as schematically shown in Fig. 1.
One-electron Hamiltonian in a magnetic field can in general be
described by
\begin{equation}
\label{Hamiltonian} H=
-\frac{1}{2m}(\vec{p}-\frac{e}{c}\vec{A})^{2}+ V_1+V_2
\end{equation}
where $V_1$ and $V_2$ are the potential created by molecules {\bf 1}
and {\bf 2}, respectively. $\vec A$ is the vector potential due to
magnetic field $\vec B$. For the simplicity and clarity, we shall
assume that the two molecules are aligned along x-direction, the
field is along the z-direction (pointing out of the paper). 
The important quantity for electron transport is the tunneling 
matrix element between two molecules. When an electron tunnels 
from an initially occupied state, say $\psi_1$ of molecule 
$\bf 1$, to empty state $\psi_2$ of molecule $\bf 2$ with 
tunneling matrix $t$ , it will contribute to the hopping 
probability $P$ (per unit time), proportional to $|t|^2\exp
(-\Delta\epsilon_{12}/(KT))$, where $\Delta \epsilon_{12}$
describes the relative energy level with respect to the Fermi
level\cite{boris,bottger}. The hopping conduction can be regarded
as an electron diffusion process in which an electron undergoes a
Brownian motion from one molecule to another, and the diffusion
constant $D$ relates to $P$ as $D=Pd^{2}$, where $d$ should be
regarded as the average distance between two neighboring molecules.
According to the Einstein relation, the electron mobility $\mu$ is
given by $\mu=eD/(KT)$ which is related to the conductivity in the
conventional way\cite{bottger}. Therefore, we can concentrate on how
the tunneling matrix element depends on the magnetic field in order
to study the magnetoresistance of the system.

In the tight-binding approximation\cite{landau}, one of the authors
in an early publication\cite{xrw} has generalized the Bardeen's
transfer matrix formalism to high dimension and in the presence of a
magnetic field. In 3D, it is
\begin{equation}
\label{hopping} t=\frac{\hbar^2}{m}\int [ (\psi^\star_1 \frac
{\partial \psi_{2}}{\partial x} - \psi_{2}\frac
{\partial\psi^\star_1} {\partial x}) -
\frac{2i}{\phi_0}(\vec{A}\cdot \hat{x})
\psi^{\star}_{1}\psi_{2}]|_{x=\frac{d}{2}}dydz,
\end{equation}
where $\phi_0=c\hbar /e$ is the flux quanta. The integration is over
the plane of $x=d/2$. For small $\vec{A}$ when the magnetic length
$l_B=\sqrt{\phi_0/(\pi B)}$ is bigger than $d$, magnetic confinement
that is responsible for the exponential increase of resistance in
the usual hopping conduction can be neglected and $\psi_1$ and
$\psi_2$ do not depends on $B$ to the zero order approximation. 
Then the magnitude of the field-independent part of $t$ is 
order of $$\frac{\hbar^2} {m\xi}\int\psi^\star_1\psi_2|_{x=
\frac{d}{2}}dydz$$ while that of the field dependent part 
is about $$\frac{\hbar^2}{m}\frac {d}{l_B^2}\int \psi^\star_1
\psi_2|_{x=\frac{d}{2}} dydz.$$ $\xi^{-1}=\xi_2^{-1}-
\xi_1^{-1}$, and $\xi_1$ and $\xi_2$ are the localization 
lengths of wavefunction $\psi_1$ and $\psi_2$, respectively. 
The two terms are comparable if $\xi d/l_B^2$ is not too small. 
For an organic semiconductor with $d=20nm$ and $\xi\simeq d$ 
in a field of $B=10mT$, $d/l_B$ is about $0.1$. Thus one shall 
expect an increase of $t$ in the field by $1\%$, same order of 
experimentally observed OMFE. For $B=100mT$, $d/l_B$ is about 
$0.33$ and $t$ increases by $10\%$! In reality, $\xi$ should be 
much bigger than $d$, especially when the hopping involves higher 
excited states as it is the case in photon-involved processes. 
Then the field-induced hopping coefficient could be even 
bigger than above estimated value, resulting in even bigger OMFE. 

Due to the Van der Waals bonding, the electric and optical signals
of organic semiconductors are too small to be detected without the
illumination of a light or applying a high electric field. 
This is why the OMFE are measured under an optical injection of 
carriers (photoluminescence and photocurrent measurements) or an 
electric field above a threshold (electric injection current and
electroluminescence measurements). When an organic semiconductor is
under the illumination of a light or under a high electric field,
the field dependent $t$ will also lead to a field dependence of
polaron density. Take optical injection of carriers as an example,
under the illumination of a light, an electron in a highest occupied
molecular orbit (HOMO) absorbs a photon and jumps to a higher empty
molecular orbit of the same molecule. As schematically illustrated
in Fig. 2, the excited electron can either dump its excessive
kinetic energy to the other degrees of freedom of the system and
form an exciton with the hole left behind or jumps to neighboring
molecules and become polarons. Depending on the relative
probabilities of excited electrons (holes) staying in the same
molecules and jumping to different molecules, the polaron 
density shall vary with the illumination intensity. Let us denote
the probability (per unit time) of a pair of electron and hole
forming an exciton in the same molecule by $P_0\sim \hbar/\tau$,
where $\tau$ is the typical time for a pair of electron and hole to
form an exciton. $P_0$ is not sensitive to a weak field since the
field cannot change much molecule orbits that determine $P_0$.
Then the polaron generation rate per unit volume is $JP/(P_0+P)$
where $J$ is the photon absorption rate per unit volume
and $P\propto |t|^2$ is the intermolecular hopping probability.
Without the illumination of a light, polaron density shall reach 
its equilibrium density $n_0'$ at a rate of $\gamma (n-n_0')$, 
where $\gamma$ is polaron decay rate. At balance, $JP/(P_0+P)=
\gamma (n-n_0')$, thus the photon-generated polaron density 
$n$ should be $n_0'+JP/[\gamma(P_0+P)]$. Clearly, $B$-dependence 
of $P$ results in a $B-$dependence of polaron density.
\begin{figure}[tbph]
\begin{center}
\includegraphics[width=0.9\columnwidth]{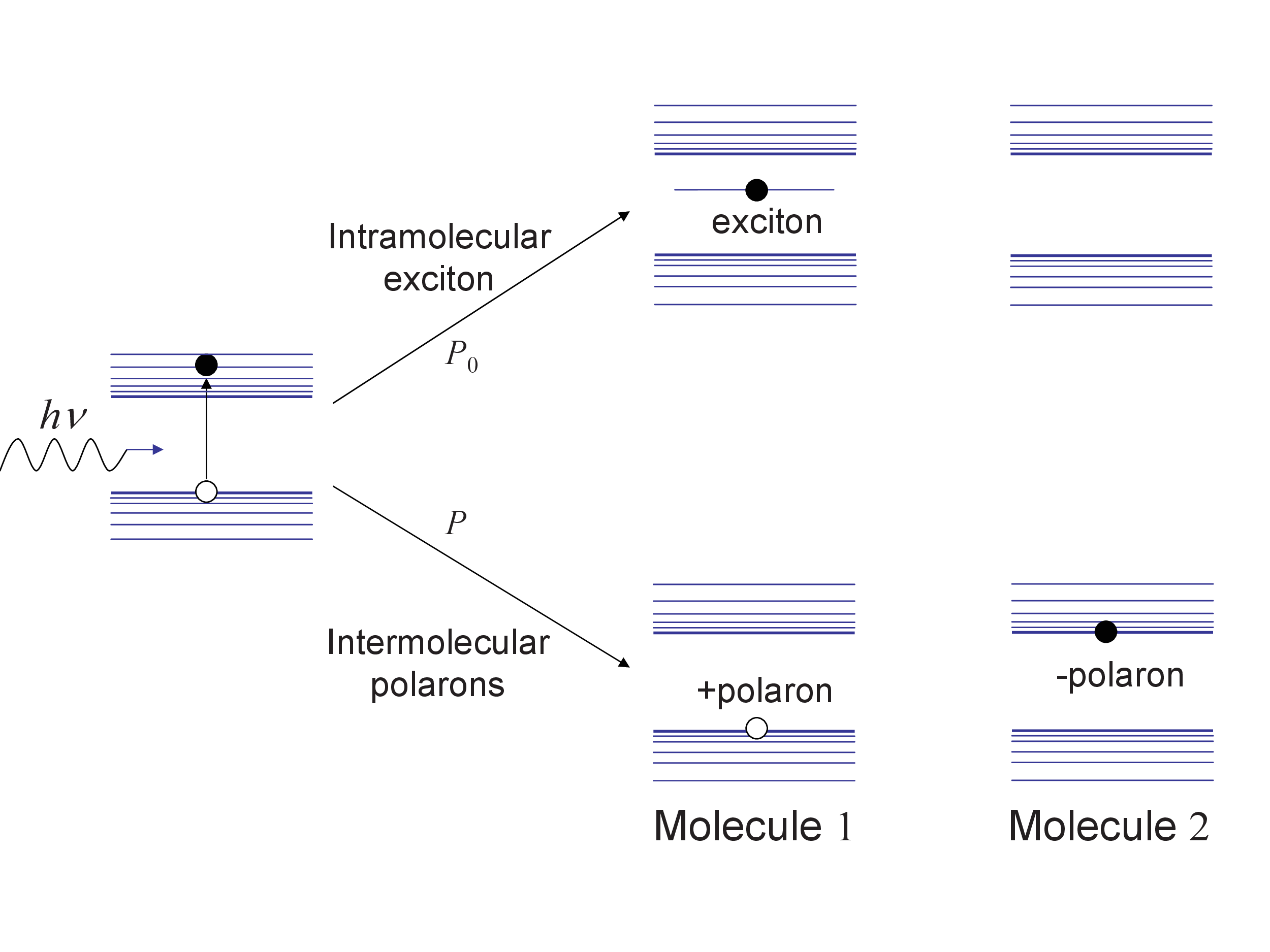}
\end{center}
\caption{Schematic illustration of polaron and exciton formation
after a pair of electron and hole is created by a photon absorption.
The excited electron-hole pair has probability $P$ jumping to the 
neighboring molecules to form positive-charged and negative-charged 
polarons, and probability $P_0$ to form an exciton. 
} \label{fig2}
\end{figure}

It is an experimental fact that the OMFE can often be fitted by
two empirical functions $B^2/(B^2+B_0^2)$ and $[B/(B+B_0)]^2$
\cite{wohlgenannt1}. A correct theory should be able to explain
why this is the case. A natural question is whether the present
picture can provide a base for these functions. According to Eq.
\eqref{hopping}, $t$ take a form of $B_0+iaB$ with $B_0$ and
$a$ real and field-independent parameters if $\psi_1$ and $\psi_2$
are real functions. This is the case when the molecule orbits
involved in hopping are localized or not degenerated\cite{landau}.
In this case, $P\propto|t|^2=a(B^2+B_0^2)$ and the polaron
density shall depend on the magnetic field as 
$\frac{P}{\gamma(P_0+P)}J+n_0'=n_0+\alpha B^2/(B^2+B_0^2)$, 
where $n_0$, $n_0'$, $\alpha$ and $B_0$ are B-independent
parameters that depend on the molecule orbits involved.
Thus, $B^2/(B^2+B_0^2)$ is a natural OMFE function for 
$t=B_0+iaB$. Interestingly, this function also appears in 
the magnetization expectation value involving the hyperfine
interactions when only quantum spin precession is considered
and all other processes are neglected \cite{wohlgenannt2}.
However, it is puzzle to us how one relates z-component
spin to the resistance and how the spin polarization can
survive under the huge thermal interaction. Surprisingly,
the second type of empirical function can also naturally
appear in when $t$ takes a form of $i(B_0+aB)$.
According to Eq. \eqref{hopping}, this can happen when the 
spatial derivatives of $\psi_1$ or $\psi_2$ are the 
functions multiplied by pure imaginary numbers. 
Of course, this must correspond to degenerated states.
In this case, the leading term in the polaron density takes a
form of $[B/(B+B_0)]^2$ in a similar argument when $P\gg P_0$.
In reality, electron (polaron) hopping between two organic
molecules should involve many molecule orbits, especially in
photophysical processes and in a high electric field.
One then needs to add contributions from all hopping events.
Thus, it is likely that both $B^2/(B^2+B_0^2)$ and
$[B/(B+B_0)]^2$ processes are presented, and OMFE should
then be fitted by the linear combinations of these
two functions, consistent with experimental findings.

The novel mechanism is very robust against the temperature and
other variations. At the room temperature, the transport of
charge carriers will involve many different molecule orbits.
Each hopping event will subject to the influence of this mechanism
as long as magnetic confinement is negligible ($l_B>d$) and $d\xi
/l_B^2$ is not too small (order of 1). Of course, one needs to
take thermal average over all hopping events. In real experiments,
organic semiconductors are highly unlikely to have well defined
crystal structures due to the nature of organic molecules.
It is then reasonable to assume that molecule-molecule orientation
of samples with a large number of molecules is quasi-random, 
meaning isotropic at large length scale, and the magnetic field can 
be along any direction with respect to the molecule-molecule bond  
instead of perpendicular direction as assumed in above discussion.
This explains why OMFE is not sensitive to the field direction in
devices. According to Eq. \eqref{hopping}, different angle between
the field and molecule-molecule bond leads to different hopping
coefficient. It should also be emphasized that the mechanism
present here does not depend on electron spins, and it does
not require large energy splits of different spin configurations.
Obviously, the picture is equally applicable to both bipolar and
hole-only (or electron-only) devices.
Differ from the previous theories that try to relate the OMFE 
with the electronic structure (either charge or spin state)
changes, the present theory attributes the OMFE to the change 
of electron hopping coefficient in a field. 
Thus, it does not have all the troubles as those spin-dynamics
related theories involve concepts of excitons and bipolarons
\cite{Kalinowski,Prigodin,Hu2,Robbert,Majumdar,Sheng,Rybicki}.

The strong OMFE in nonmagnetic organic materials is the consequences
of combined effects of the Van der Waals bonding between organic
molecules, a large molecule-molecule separation and localization
length. In the usual inorganic semiconductor, all these conditions
are not satisfied. Firstly, the bonding between atoms are covalent
so that the field-independent hopping coefficient is order of
several eV, much bigger than the field-related contribution.
Secondly, in the hopping conduction of the conventional doped
semiconductors, the localization length of a localized state and 
the hopping distance are order of angstroms so that $\xi d/l_B^2$ 
is many orders magnitude smaller than that in the organic
semiconductors in the weak field. In a strong field $l_B<d$, the 
opposite regime of OMFE phenomenon, magnetic confinement dominates 
electron hopping, and the resistance increases exponentially 
with the magnetic field. This is why the similar magnetic-field 
effects had not been observed in inorganic semiconductors. 
Whether the mechanism presented here is genuine or not is subject 
to the experimental tests. Thus, we are proposing to design and 
manufacture devices with various $\xi d/l_B^2$ by using different 
materials and molecule structures. It would be a definite proof 
of present theory if all devices show OMFE when $d/l_B<1$ and
$\xi d/l_B^2$ is order of 1.

In conclusion, we present a novel mechanism for the OMFE for
nonmagnetic organic semiconductors. The mechanism is very general
and robust for organic semiconductors, but is normally not important
for usual covalently bonded inorganic semiconductors. The mechanism
can not only explain all experimentally observed OMFE, but also
provide a solid theoretical bases for the empirical OMFE formulas.
New experiments are needed to firmly establish this mechanism as the
genuine cause of the OMFE.

This work is supported by Hong Kong UGC grants (\#604109,
HKUST17/CRF/08, and RPC07/08.SC03). S.J. Xie is supported by the
National Basic Research Program of China (Grant No.2009CB929204 and
No.2010CB923402) and the National Natural Science Foundation of the
People's Republic of China (Grant No. 10874100)

\end{document}